# Exceptional points and lines and Dirac points and lines in magnetoactive cholesteric liquid crystals

A.H. Gevorgyan

Far Eastern Federal University, Institute of High Technologies and Advanced Materials, 10 Ajax Bay, Russky Island, Vladivostok 690922, Russia

We investigated the properties of cholesteric liquid crystals (CLCs) being in external static magnetic field directed along the helix axis. We have shown that in the case of the wavelength dependence of magneto-optic activity parameter, and in the presence of absorption new features appear in the optics of CLCs. We have shown that in this case new photonic band gaps (PBGs) appear. This new PBG is sensitive to the polarization of the incident light. But if the chirality sign of the polarization of the incident diffracting light for the basic PBG (which exist also at the absence of external magnetic field) is determined only by the chirality sign of the CLC helix, then for the second one it is determined by the external magnetic field direction (i.e., on whether the directions of the external magnetic field and the incident light are parallel, or they are antiparallel). We have shown that in this case besides Dirac points there appear also Dirac lines as well as exceptional points and exceptional lines. And moreover, at some of these points and lines there appear the lines or wide bands of magnetically induced transparency, on others a wide coherent perfect absorption band appears that is insensitive to incident light polarization. And finally on some points the same reflection, transmission and absorption takes place for any polarization of incident light. This system can be applied as tunable narrow-band or broad-band filters and mirrors, a highly tunable broad/narrow-band coherent perfect absorber, transmitter, ideal optical diode, and other devices.

## 1. Introduction

Many physical phenomena firstly originated in quantum physics have found important and useful realizations in optical systems. Chiral bound states (Fano resonant states with long optical lifetimes controlled by symmetry-breaking perturbations) [1,2], Dirac points (the points, where the intersection of any two wave vector curves takes place (they degenerate) and a linear law of the wave vector dependence on the frequency near these points takes place, too) [3-5], topological insulators (materials whose interior behaves as an electrical insulator while their surface behaves as an electrical conductor, meaning that electrons can only move along the surface of the material) [6-9], bound states in the continuum ( waves that remain localized even though they coexist with a continuous spectrum of radiating waves that can carry energy away) [10-12], Bloch oscillations (oscillations of electrons within the Brillouin zone induced by an applied electric field) [13], Zener tunneling (the direct tunneling of a Bloch particle into a continuum of another energy band, which takes place without extra energy in the presence of a large electric field in the crystal) [14], electromagnetically induced transparency (EIT; which is a quantum interference effect in three-level atomic systems that eliminates the absorption at the resonance frequency and gives rise to a narrow transparency window) [15,16], electromagnetically induced absorption (EIA; which is the counter phenomenon of invoking constructive interference between multiple interaction pathways to enhance/induce absorption) [17] and etc., are examples of effects that have opened rich lines of research in photonics. They are of continuously growing interest for both the fundamental understanding of wave phenomena and its application to photonic devices. Occurrence of exceptional points is another salient example [18-27]. As mentioned, a Dirac point appears when two bands cross each other locally and exhibit a linear dispersion in any direction in the momentum space. As the eigenvalues of Hermitian systems are real, two orthogonal eigenstates coexist at the Dirac point with the same eigenvalue. The counterpart in non-Hermitian systems are exceptional points, where the complex eigenvalues of two different bands are identical, with both equal real and imaginary parts. In exceptional points, the eigenvectors and therefore the bands are also degenerate. The non-Hermitian photonic structures can be created by introducing absorption or gain in Hermitian photonic structures.

In recent years, the possibilities of observing the above-mentioned quantum effects analogs in a wide variety of photonic structures and the application of these effects have been intensively investigated. Moreover, some photonic structures in external magnetic field or nanostructures with strong magnetic dipole interactions exhibit a narrow transparency window, and some others enhance/induce absorption. These new phenomena by analogy of EIT and EIA are called magnetically induced transparency (MIT), and magnetically induced absorption (MIA), respectively [28-39]. In connection with the above effects, of particular interest are cholesteric liquid crystals (CLCs). CLCs are 1D photonic structures with rich optical properties and which find a wide variety of applications in a wide variety of fields (see, in particular, [40] and references therein). Magneto-optics of cholesteric liquid crystals (CLCs) is of particular interest. In the papers [41-50] the results of a theoretical and experimental study of the magneto-optical properties of CLCs are presented. Moreover, in the papers [51-54] the existence of MIT and MIA was demonstrated. In [55] it was reported about observation of Dirac points in CLCs. It was shown that the MIT and MIA phenomena are observed near the Dirac points. Let us note, that all the works [51-55] deal with the case when the parameter of magneto-optical activity g is constant and is independent of wavelength.

In this paper, we will discuss the possibility of observing Dirac points and lines and exceptional points and lines in magnetically active CLCs in the case when the parameter of magneto-optical activity g is function of wavelength. These effects were to be expected because, as shown in [44], in the presence of an external magnetic field directed along the axis of the helix, the wave vectors of the eigenwaves of the CLC shift not only along the wavelength axis, but also perpendicular to this axis. Two of the four wave vectors $k_m(\lambda)$ ($m=1,2,3,4$) are shifted upward parallel to the $k_m =0$ axis, while the other two are shifted downward. This means that at certain values of the parameters of the medium and the external magnetic field, the curves of the wave vectors can intersect, and Dirac points may appear. In the presence of absorption, an exceptional point may arise. Moreover, at a suitable choice of parameters of the medium and the external magnetic field it is possible to achieve not only their intersection, but also their almost exact coincidence in a sufficiently wide frequency range and the appearance of the Dirac line instead of the Dirac point, and in the presence of absorption also the appearance of the exceptional line.

## 2. Models and methodology. Results

The dielectric permittivity and magnetic permeability tensors of the magnetoactive CLC in the external magnetic field directed along the helix axis have the forms:

$$\hat{\varepsilon}(z) = \varepsilon_m \begin{pmatrix} 1 + \delta\cos2az & \pm\delta\sin2az \pm ig/\varepsilon_m & 0 \\ \pm\delta\sin2az \mp ig/\varepsilon_m & 1 - \delta\cos2az & 0 \\ 0 & 0 & 1 - \delta \end{pmatrix}, \text{ and } \hat{\mu}(z) = \hat{I}, \quad (1)$$

where g is the parameter of magneto-optical activity of CLC, and it, in general, is a function of the external magnetic field, Verdet constant, and dielectric permittivity of media, $\varepsilon_m = (\varepsilon_1 + \varepsilon_2)/2$, $\delta = \frac{(\varepsilon_1-\varepsilon_2)}{(\varepsilon_1+\varepsilon_2)}$, $a = 2\pi / p$, $p$ is the pitch of the helix, axis $z$ and external magnetic field are directed along CLC helix axis. We take the dependence of g on wavelength as follows [56]:

$$g = \frac{VB_{ext}\lambda}{\pi}\sqrt{\left|\varepsilon_m - \left(\frac{VB_{ext}\lambda}{2\pi}\right)^2\right|}, \quad (2)$$

where $V$ is the Verdet constant, and $B_{ext}$ is the external magnetic field induction and $\lambda$ is the light wavelength in vacuum. This function $g(\lambda)$ has two zeros on the wavelengths $\lambda_1 = 0$ and $\lambda_2 = \frac{2\pi}{VB_{ext}}\sqrt{\varepsilon_m}$, and passes across local maximum on the wavelength $\lambda_3 = \frac{\pi}{VB_{ext}}\sqrt{2\varepsilon_m}$. $\lambda_3$ is between $\lambda_1$ and $\lambda_2$. After $\lambda_2$, g increases monotonically.

We consider the case of light propagation along the helix axis (along the z-axis). Let as note that the exact analytical solution of the Maxwell's equations for a CLC when light propagates along its axis at the absence of external magnetic field is known [57, 58]. In this case, passing to a rotating coordinate system we will seek the solutions of Maxwell's equations in the form [45]:

$$\vec{\mathcal{E}}(z,t) = \sum_{m=1}^{4} \vec{\mathcal{E}}_{0m} \exp(ik_{mz}z) \exp(-i\omega t), \qquad (3)$$

where $k_{mz}$ are the z components of wave vectors in the rotating coordinate system. The electric field in the laboratory coordinate system has the form:

$$\vec{E}(z) = \hat{R}^{-1}(az)\vec{\mathcal{E}}(z), \qquad (4)$$

where $\hat{R}(az) = \begin{pmatrix} \cos az & \mp \sin az & 0 \\ \pm \sin az & \cos az & 0 \\ 0 & 0 & 1 \end{pmatrix}$ is the rotation matrix.

Substituting (3) into Maxwell equations we obtain the following dispersion equation:

$$\left(\frac{\omega^2}{c^2}\varepsilon_1 - k_{mz}^2 - a^2\right)\left(\frac{\omega^2}{c^2}\varepsilon_2 - k_{mz}^2 - a^2\right) - \left(2ak_{mz} - \frac{\omega^2}{c^2}g\right)^2 = 0. \qquad (5)$$

Now, we can solve the problem of light reflection, transmission, and absorption in the case of a planar magnetoactive CLC layer of finite thickness. We assume that the optical axis of this CLC layer is perpendicular to the boundaries of the layer and is directed along the z-axis. The CLC layer on its both sides border with isotropic half-spaces with the same refractive indices equal to $n_s$. The boundary conditions, consisting of the continuity of the tangential components of the electric and magnetic fields, are a system of eight linear equations with eight unknowns (in more details see [41]). Solving this boundary-value problem, one can determine the values of the reflected $\mathbf{E}_r$ and transmitted $\mathbf{E}_t$ fields and calculate the energy coefficient of reflection $R = \frac{|\mathbf{E}_r|^2}{|\mathbf{E}_i|^2}$, transmission $T = \frac{|\mathbf{E}_t|^2}{|\mathbf{E}_i|^2}$, and absorption $A = 1 - (R + T)$, where $\mathbf{E}_i$ is the incident light field. Here and below, we will consider the case of minimal influence of dielectric boundaries, that is the case $n_s = \sqrt{\varepsilon_m}$. CLCs have a self-organized helical structure (it can be regarded as a 1D PC), the periodic structure of which gives rise to a polarization-sensitive photonic band gap (PBG). The PBG of the CLCs is defined by the condition $pn_o < \lambda < pn_e$, where $n_o = \sqrt{\varepsilon_2}$ and $n_e = \sqrt{\varepsilon_1}$ are the ordinary and extraordinary local refractive indices of CLC. Within the PBG only the circularly polarized light having the same handedness as the CLC helix is selectively reflected (of course at normal light incidence). Below, we call this PBG as the first (or basic) PBG. The helix of our CLC is right-handed, so the right circular polarization is diffracting, and left one non diffracting.

Note that, below, in our plots, we will present the dependences of the real and imaginary parts of the wave numbers on frequency, in order to demonstrate the existence of a linear or nonlinear dependence of these quantities on frequency, in particular near the Dirac points and the Dirac lines, as well as near the exceptional points and the exceptional lines, while for reflection, transmission and absorption, we will represent their spectra, i.e. the dependence of these quantities on the wavelength.

We will mainly investigate the features of reflection (transmission and absorption) spectra for incident light with polarizations coinciding with eigenpolarizations (EPs). The EPs are the two polarizations of incident light that do not change as the light passes through the system. These two EPs for CLC approximately coincide with orthogonal circular polarizations. Some discrepancy arises only near the photonic band gaps (PBGs). Let us now enumerate the EPs in the following way, we will assume that the first EP is the EP that approximately coincides with the right-hand circular polarization, that is with diffracting circular polarization, while the second EP with the left-hand one.

Fig. 1 shows the reflection, transmission and absorption spectra at different values of helix pitch at the presence of anisotropic absorption. The incident light has polarization coinciding with the first EP ($R_1, T_1, A_1$) and with the second EP ($R_2, T_2, A_2$). In Fig. 1 (a, c, e, g, i, k) the directions of the incident light and the external magnetic field coincide, while in Fig. 1 (b, d, f, h, j, l) they are oppositely directed.

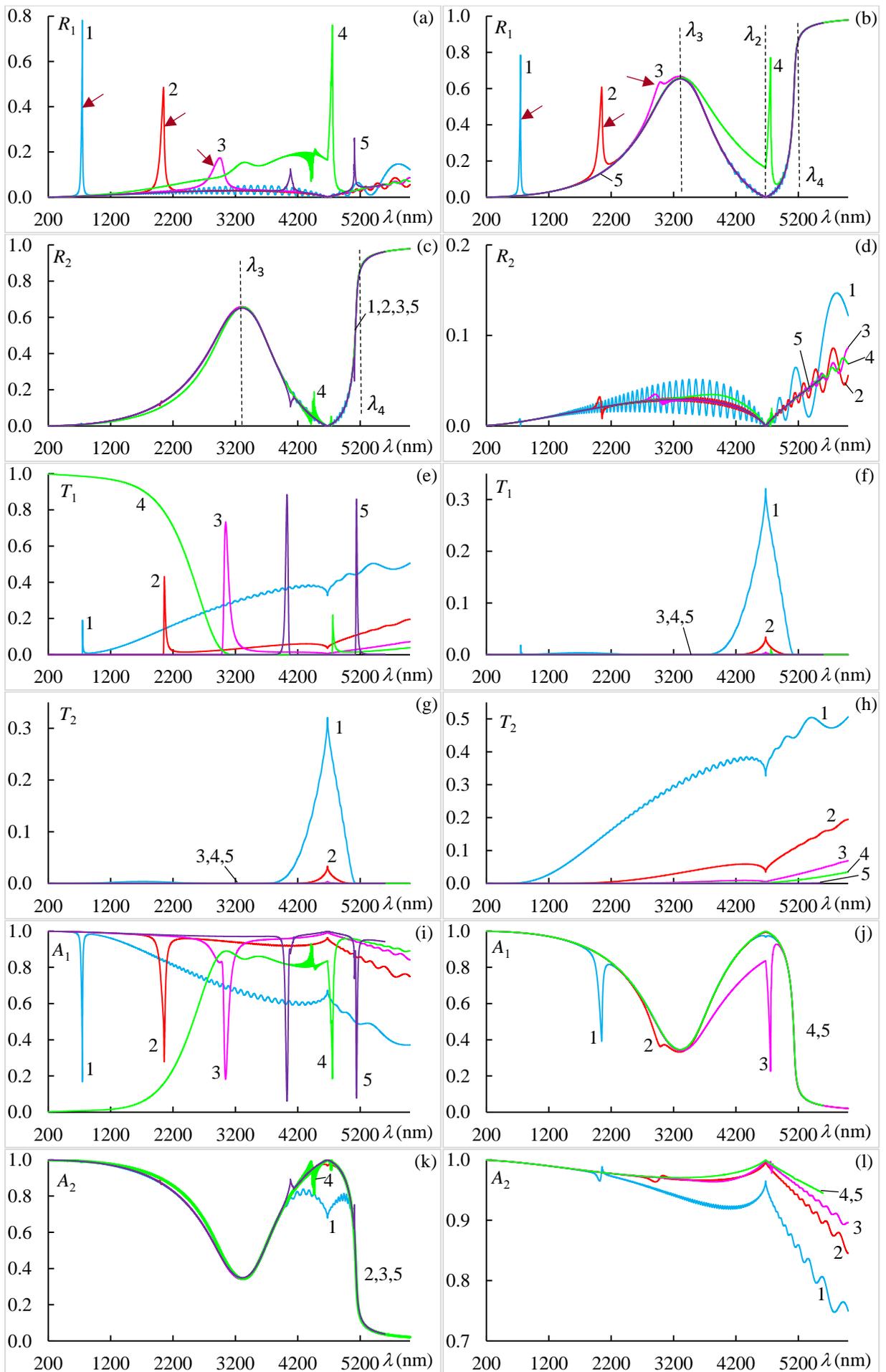

Fig. 1. The spectra of reflection, transmission and absorption at different values of helix pitch. The incident light has polarization coinciding with the first EP ($R_1, T_1, A_1$) and with the second EP ($R_2, T_2, A_2$). In Fig. 1 (a, c, e, g, i, k) the directions of the incident light and the external magnetic field coincide, while in Fig. 1 (b, d, f, h, j, l) they are oppositely directed. $p = 500$ nm (curves 1), $p = 1500$ nm (curves 2), $p = 2500$ nm (curves 3), $p = 3200$ nm (curves 4) and finally $p = 5500$ nm (curves 5). $\text{Re}\varepsilon_1 = 2.29, \text{Im}\varepsilon_1 = 0., \text{Re}\varepsilon_2 = 2,143, \text{Im}\varepsilon_2 = 0.1, d=50p$.

As is well known (see, in particular, [44]), an external magnetic field leads to a displacement of the first PBG, and both at g> 0 and at g<0 this displacement is of the same order of magnitude and is directed towards short wavelengths. Next, as can be seen from Fig. 1 (a,b), with an increase in the pitch of the helix, the height of the peaks of reflection of the first PBG decreases (in Fig. 1 a, b, these peaks are highlighted by brown arrows). Moreover, a new PBG is formed near the wavelength $\lambda_3$ (Fig. 1 b,c). We call this PBG the second PBG. This new PBG is sensitive to the polarization of the incident light. But if the chirality sign of the polarization of the incident diffracting light for the first PBG is determined only by the chirality sign of the CLC helix, then for the second one it is determined by the external magnetic field direction (i.e., on whether the directions of the external magnetic field and the incident light are parallel, or they are antiparallel). In the first case, the diffraction reflection undergoes the light with the second EP, and in the second case, the light with the first EP.

In order to answer the question whether the wavelength $\lambda_3$ is a Dirac point or an exceptional point, we will investigate the dependence of the real and imaginary parts of the wave numbers on frequency. Equation (5) is a fourth-degree equation with respect to wavenumbers $k_{mz}$ and therefore has four roots (m=1,2,3,4). In the absence of absorption and external magnetic field there is a frequency band (with the boundaries $\omega_{01} = \frac{ca}{\sqrt{\varepsilon_1}}$ and $\omega_{02} = \frac{ca}{\sqrt{\varepsilon_2}}$) where two wave numbers out of four are purely imaginary ($\text{Im}k_{mz} \neq 0$ and $\text{Re}k_{mz} = 0$), and we will call them resonant wave numbers to differentiate from the two others which we will call non-resonant ones. Namely here the first PBG exists. Now let us enumerate the eigensolutions of Eq. (5) in the following way: $m = 1$ and 4, correspondingly, for the non-resonance wave vectors, and $m = 2$ and 3, correspondingly, for the resonance wave vectors.

Fig. 2 shows the dependences of $\text{Re}k_{mz}$ (a) and $\text{Im}k_{mz}$ (b,c,d) on the frequency $\omega$ in the presence of external magnetic field, but in the absence of absorption. In the presence of an external magnetic field, instead of the first direct PBG, we have an indirect PBG (in this PBG the real parts of the resonant wave numbers are different from zero, see band between $\omega_{01}$ and $\omega_{02}$), and it experiences a blueshift, as mentioned above. As can be seen from Fig. 2, near the point $\omega_3 = \frac{2\pi c}{\lambda_3}$ we have the touching of wave numbers $\text{Re}k_{1z}$ and $\text{Re}k_{2z}$, and they depend on $\omega$ linearly near this point. Since near this point we have $\text{Im}k_{1z} \neq \text{Im}k_{2z}$, and, moreover, both in the presence of isotropic absorption (at $\text{Im}\varepsilon_1 = \text{Im}\varepsilon_2$) and in the presence of anisotropic absorption (at $\text{Im}\varepsilon_1 \neq \text{Im}\varepsilon_2$), then here we have exactly the Dirac point and a second PBG formed near this point (see Figs. 1b and c). At $|g| \geq \varepsilon$ two out of four wave numbers become complex at the absence of absorption and a new (the third) PBG will appears in the region $\omega \leq \omega_4 = \frac{\sqrt{2}VB_{ext}c}{\sqrt{\varepsilon_m(1+\sqrt{2})}}$ (again see, also Fig.1 b and c), that is this new PBG is observed both at g > 0 and at g < 0. Let us note, that in the case of $|g| \geq \varepsilon$ at the absence of absorption two from four wave numbers of homogeneous magnetically active media also become complex.

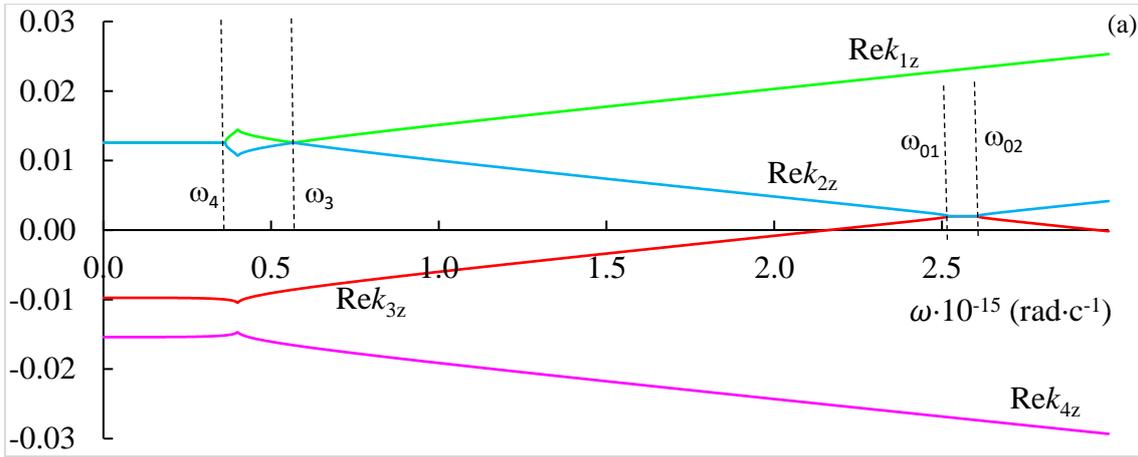

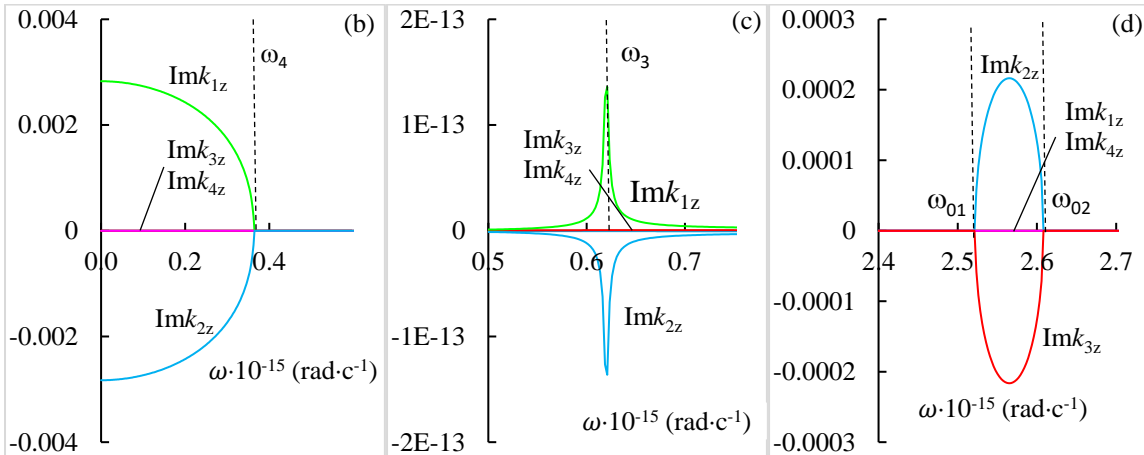

Fig. 2. The dependences of $\text{Re}k_{mz}$ (a) and $\text{Im}k_{mz}$ (b,c,d) on the frequency $\omega$ in the presence of external magnetic field, but in the absence of absorption. The parameters are: $p = 500$ mn, $\text{Re}\varepsilon_1 = 2.29$, $\text{Im}\varepsilon_1 = 0.$, $\text{Re}\varepsilon_2 = 2,143$, $\text{Im}\varepsilon_2 = 0$.

Then, the effect of absorption on the wave number-frequency dependencies near the $\omega_3$ point was investigated. Fig. 3 shows the dependences of $\text{Re}k_{1z}$ and $\text{Re}k_{2z}$ (a) and $\text{Im}k_{1z}$ and $\text{Im}k_{2z}$ (b) on the frequency $\omega$ in the absence of absorption (solid lines) and in the presence of isotropic absorption (dashed lines) with parameters $\text{Im}\varepsilon_1 = \text{Im}\varepsilon_2 = 0.1$. As can be seen from Fig. 3a, in the presence of absorption there is no longer touching of the curves $\text{Re}k_{1z}(\omega)$ and $\text{Re}k_{2z}(\omega)$ at this point at the frequency $\omega_3$, and as absorption increases, they move further and further away from each other at this point. At anisotropic absorption with parameters $\text{Im}\varepsilon_1 = 0.2$, $\text{Im}\varepsilon_2 = 0.$ we obtain with high accuracy the same results as at isotropic absorption with parameters $\text{Im}\varepsilon_1 = 0.1$, $\text{Im}\varepsilon_2 = 0.1$, i.e., the type of absorption practically has no influence on the dependences $\text{Re}k_{1z}(\omega)$ and $\text{Re}k_{2z}(\omega)$ near the point on the frequency $\omega_3$.

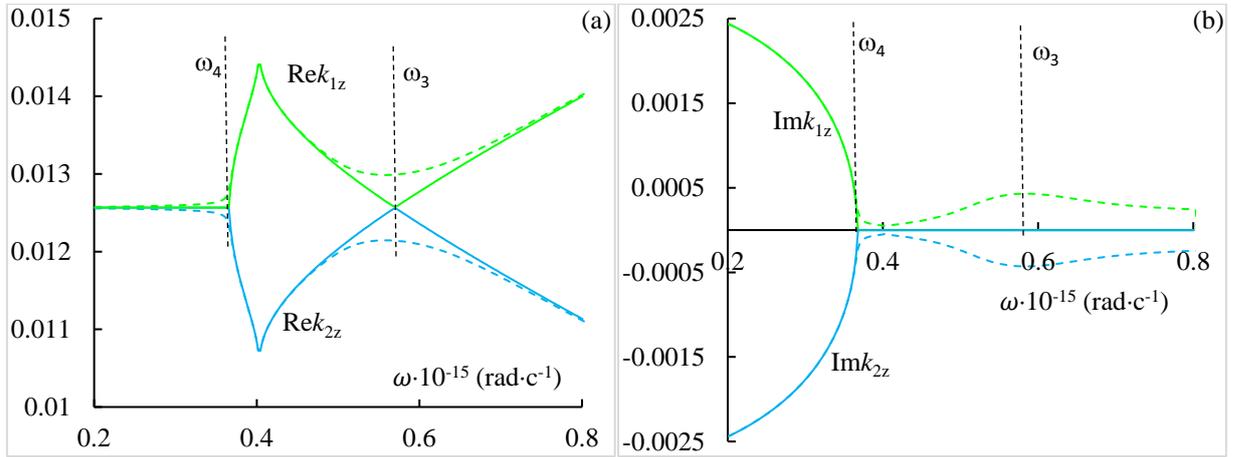

Fig. 3. The dependences of $\text{Re}k_{1z}$ and $\text{Re}k_{2z}$ (a) and $\text{Im}k_{1z}$ and $\text{Im}k_{2z}$ (b) on the frequency $\omega$ in the absence of absorption (solid lines) and in the presence of isotropic absorption (dashed lines) with parameters $\text{Im}\varepsilon_1 = \text{Im}\varepsilon_2 = 0.1$. The other parameters are the same as in Fig.2.

Once again, we return to Fig. 1. As mentioned above, the spectra on the left column in Fig. 1 correspond to the case when the directions of the incident light and the external magnetic field coincide, and the spectra on the right column to the case when these directions are opposite. These spectra are essentially different; therefore, this system is nonreciprocal and the nonreciprocity varies within wide intervals. This system is nonreciprocal both with respect to reflection and transmittance, and with respect to absorption, too.

Next, as can be seen from Fig. 1 e and i at $p = 3200$ nm, we have a broadband almost complete transmittance and, accordingly, a practical absence of light absorption in this band for incident light with the first EP in a magnetically active CLC with locally anisotropic absorption (see Fig. 1 a, e and i, curves 4), and practically full absorption of light in this band for incident light with the second EP (see Fig. 1 c, g and k, again curves 4). Again, to reveal these features, we proceed to the study of dependencies $\text{Re}k_{mz}(\omega)$ and $\text{Im}k_{mz}(\omega)$ at these parameters. But we will do this both for $p = 3150$ nm and for $p = 3200$ nm, since as show our calculations, with these values for the helix pitch we have a better manifestation of the features presented below.

Fig. 4 shows the dependences of $\text{Re}k_{mz}$ (a) and $\text{Im}k_{mz}$ (b) on the frequency $\omega$ in the presence of anisotropic absorption. As can be seen from Fig. 4, here we have a linear dependence of the real and imaginary parts $k_{mz}$ on $\omega$. It is further seen that we have a practical coincidence of the curves $\text{Re}k_{1z}$ and $\text{Re}k_{3z}$ in this frequency region. Moreover, we have also $\text{Im}k_{1z} \neq 0$, while $\text{Im}k_{3z}=0$ in this region. Hence, here we have the Dirac line.

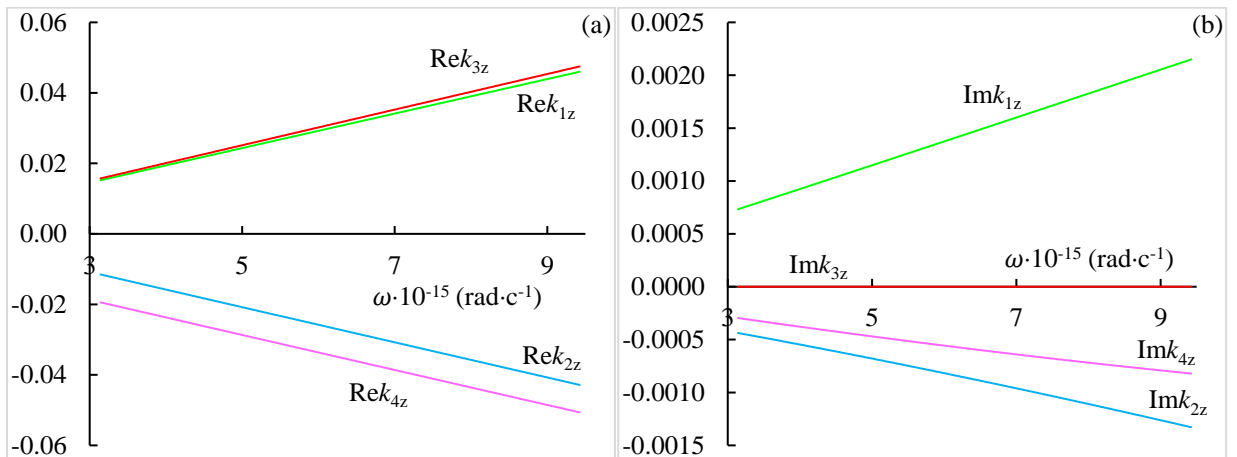

Fig. 4. The dependences of $\text{Re}k_{mz}$ (a) and $\text{Im}k_{mz}$ (b) on the frequency $\omega$ in the presence of anisotropic absorption with $\text{Im}\varepsilon_1 = 0$ and $\text{Im}\varepsilon_2 = 0.2$, $p = 3200$ nm. The other parameters are the same as in Fig.2.

Fig. 5 shows the same dependences as in Fig. 4 but in the presence of isotropic absorption. As can be seen from Fig. 5, in this case we again observe a linear dependence of the real and imaginary parts $k_{mz}$ on $\omega$ and a practical coincidence of the curves $\text{Re}k_{1z}$ and $\text{Re}k_{3z}$ in this frequency region. But, unlike the case of anisotropic absorption in this case we also have a practical coincidence of the curves $\text{Im}k_{1z}$ and $\text{Im}k_{3z}$, in this frequency region, that is we have $\text{Im}k_{1z} = \text{Im}k_{3z} \neq 0$, here. Hence, here instead of the Dirac line we have the exceptional line.

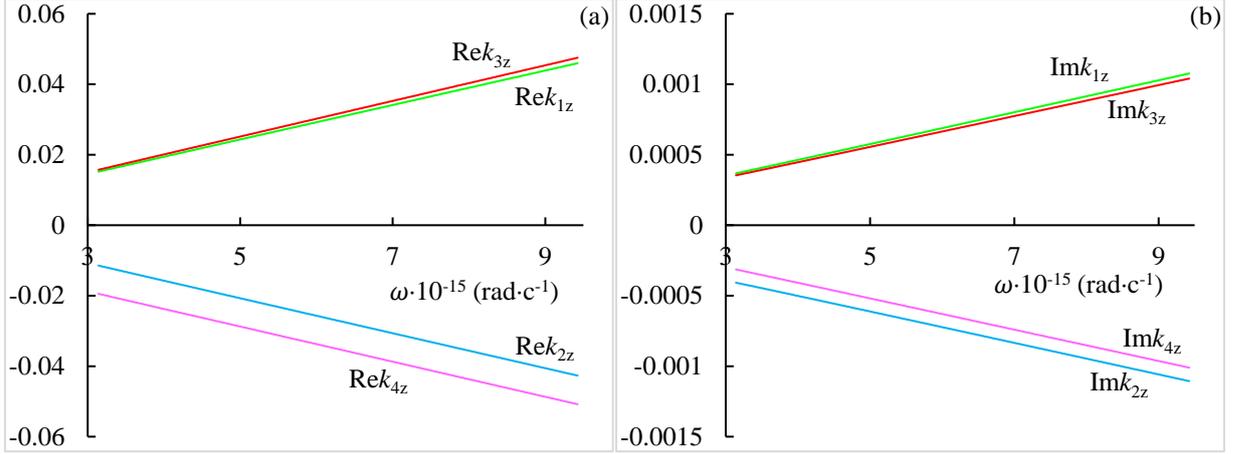

Fig. 5. The dependences of $\text{Re}k_{mz}$ (a) and $\text{Im}k_{mz}$ (b) on the frequency $\omega$ in the presence of isotropic absorption with $\text{Im}\varepsilon_1 = 0.1$ and $\text{Im}\varepsilon_2 = 0.1$. $p = 3150$ nm. The other parameters are the same as in Fig.2.

Now we investigated the reflection, transmission, and absorption spectra peculiarities in these two cases. Fig. 6 shows the spectra of reflection $R$ (curves 1,2), transmission $T$ (curves 3,4) and absorption A (curves 5,6) in the following two cases: (a) $\text{Im}\varepsilon_1 = 0.$, $\text{Im}\varepsilon_2 = 0.2$ and $p = 3200$ nm, that is in the case of anisotropic absorption, and (b) $\text{Im}\varepsilon_1 = 0.1$, $\text{Im}\varepsilon_2 = 0.1$ and $p = 3150$ nm, that is in the case of isotropic absorption. The incident light has polarization coinciding with the first EP (curves 1,3,5) and with the second EP (curves 2,4,6). As can be seen from Fig. 6 in the case of anisotropic absorption we have a wideband (in order of 2000 nm and more) total transmission for incident light with the first EP and wideband total absorption for light with the second EP. Hence, here wideband coherent perfect transmission (CPT) takes place for incident light with the first EP and wideband coherent perfect absorption (CPA) takes place for incident light with the second EP. Moreover, here this system can work as wideband unidirectional transmitter or wideband ideal optical diode. While in the case of isotropic absorption we have wideband total absorption for light with both the first and the second EPs, i.e., here we have a wide spectral region, where there is a reflection, transmission and absorption independent of the polarization of the incident light, moreover, a wide spectral region, where a total absorption takes place independent of the polarization of the incident light. Hence, here this system can work as wideband total absorber for incident light with any polarization.

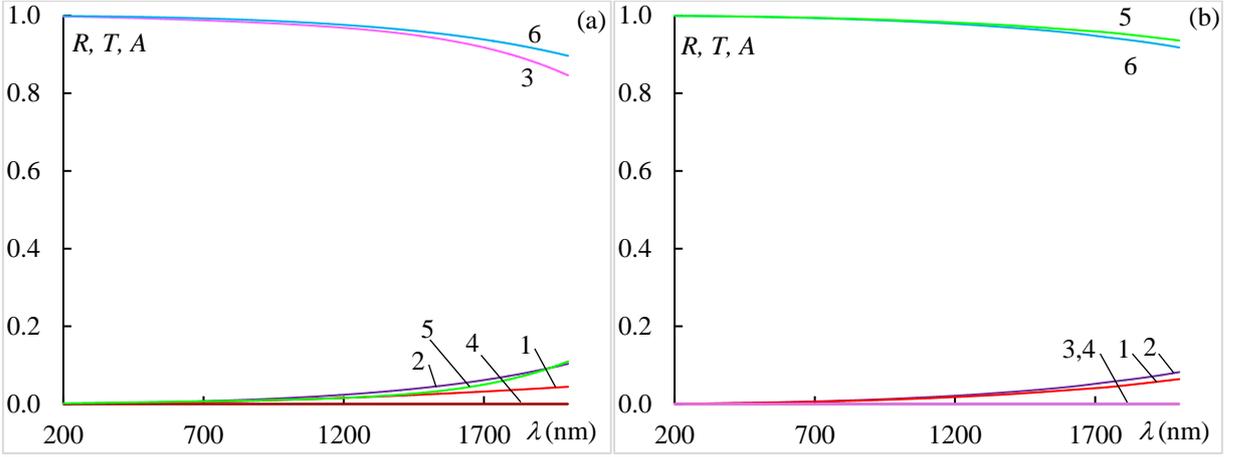

Fig. 6. The spectra of reflection $R$ (curves 1,2), transmission $T$ (curves 3,4) and absorption A (curves 5,6) in the following two cases: (a) $\mathrm{Im}\varepsilon_1 = 0.$, $\mathrm{Im}\varepsilon_2 = 0.2$ and $p = 3200$ nm, and (b) $\mathrm{Im}\varepsilon_1 = 0.1$, $\mathrm{Im}\varepsilon_2 = 0.1$ and $p = 3150$ nm. The incident light has polarization coinciding with the first EP (curves 1,3,5) and with the second EP (curves 2,4,6). $d=35p$. The other parameters are the same as in Fig.2.

We now investigate the features of Dirac points and exceptional points in the case of $p=5500$ nm values of helix pitch. First, as above, we will investigate the features of the dependences $\mathrm{Re}k_{mz}(\omega)$ and $\mathrm{Im}k_{mz}(\omega)$ in the case of both isotropic and anisotropic absorption and at absence of absorption as well. Fig. 7 shows the dependences of $\mathrm{Re}k_{mz}$ (a) and $\mathrm{Im}k_{mz}$ (b) on the frequency $\omega$ in the absence of absorption (solid lines) and in the presence of isotropic absorption with parameters $\mathrm{Im}\varepsilon_1 = \mathrm{Im}\varepsilon_2 = 0.05$ (dashed lines). As can be seen from Fig. 7a in the two points on the frequencies $\omega_5$ and $\omega_6$ the intersection or touching of curves $\mathrm{Re}k_{1z}(\omega)$ and $\mathrm{Re}k_{2z}(\omega)$ take place. Equating $\mathrm{Re}k_{1z}(\omega)$ to $\mathrm{Re}k_{2z}(\omega)$, we obtain the following approximate expressions for $\omega_5$ and $\omega_6$: $\omega_5 \approx \dfrac{c\sqrt{16\pi^4+V^4 B_{ext}^4 p^4}}{p\sqrt{\varepsilon_m(4\pi^2+V^2 B_{ext}^2 p^2)}}$ and $\omega_6 \approx \dfrac{c\sqrt{4\pi^2+V^2 B_{ext}^2 p^2}}{p\sqrt{\varepsilon_m}}$. As can be seen from Fig. 7b in these two points on the frequencies $\omega_5$ and $\omega_6$ touching of curves $\mathrm{Im}k_{1z}(\omega)$ and $\mathrm{Im}k_{2z}(\omega)$ take place, too. Therefore, these points are exceptional points.

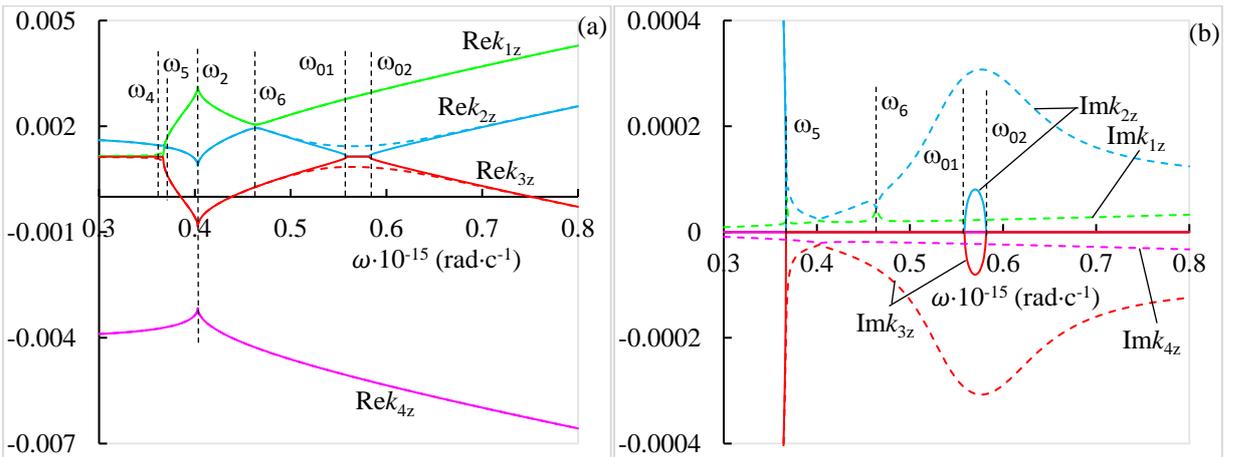

Fig. 7. The dependences of $\mathrm{Re}k_{mz}$ (a) and $\mathrm{Im}k_{mz}$ (b) on the frequency $\omega$ in the absence of absorption (solid lines) and in the presence of isotropic absorption with parameters $\mathrm{Im}\varepsilon_1 = \mathrm{Im}\varepsilon_2 = 0.05$ (dashed lines). $p = 5500$ nm. The other parameters are the same as in Fig.2.

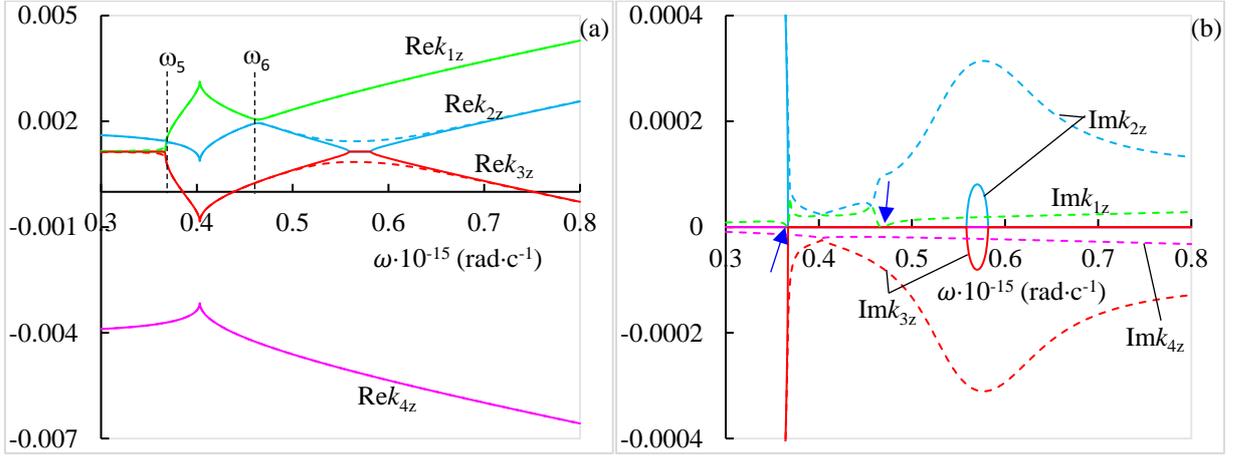

Fig. 8. The dependences of $\text{Re}k_{mz}$ (a) and $\text{Im}k_{mz}$ (b) on the frequency $\omega$ in the absence of absorption (solid lines) and in the presence of anisotropic absorption with parameters $\text{Im}\varepsilon_1 = 0$ and $\text{Im}\varepsilon_2 = 0.1$ (dashed lines). The other parameters are the same as in Fig.7.

Fig. 8 shows the dependences of $\text{Re}k_{mz}$ (a) and $\text{Im}k_{mz}$ (b) on the frequency $\omega$ in the absence of absorption (solid lines) and in the presence of anisotropic absorption with parameters $\text{Im}\varepsilon_1 = 0$ and $\text{Im}\varepsilon_2 = 0.1$ (dashed lines). As can be seen from Fig. 8a again in these two points on the frequencies $\omega_5$ and $\omega_6$ the intersection or touching of curves $\text{Re}k_{1z}(\omega)$ and $\text{Re}k_{2z}(\omega)$ take place. And, as can be seen from Fig. 8b in these two points on the frequencies $\omega_5$ and $\omega_6$ touching of curves $\text{Im}k_{1z}(\omega)$ and $\text{Im}k_{2z}(\omega)$ take place, too. Therefore, these points again are exceptional points. But one more new feature is also observed here, namely, near these exceptional points at the frequencies $\omega_5$ and $\omega_6$ there are two points where $\text{Im}k_{1z}(\omega) = 0$ (in Fig. 8b these points are indicated by blue arrows).

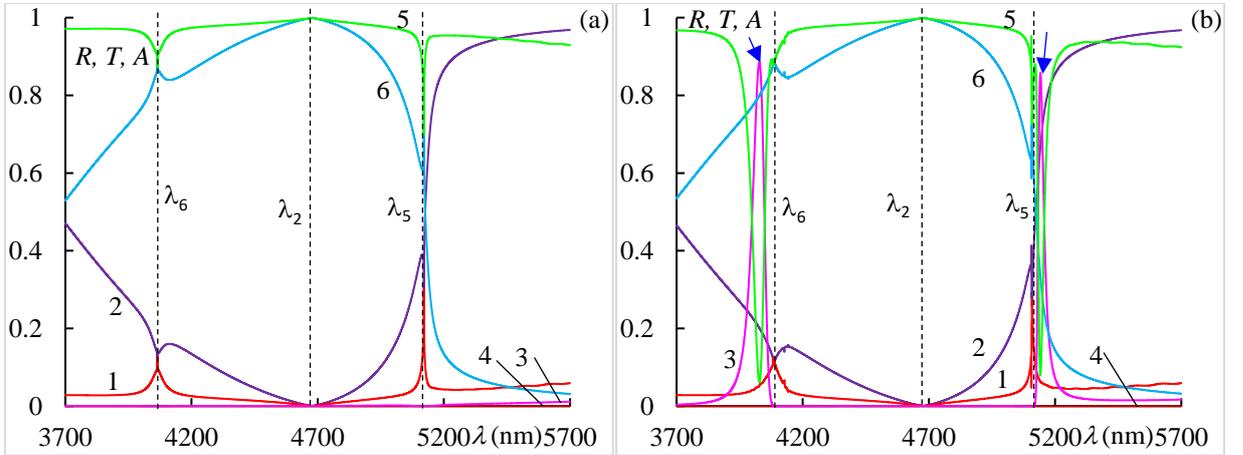

Fig. 9. The spectra of reflection $R$ (curves 1,2), transmission $T$ (curves 3,4) and absorption A (curves 5,6) in the following two cases: (a) $\text{Im}\varepsilon_1 = 0.05$, $\text{Im}\varepsilon_2 = 0.05$, and (b) $\text{Im}\varepsilon_1 = 0.$, $\text{Im}\varepsilon_2 = 0.1$. The incident light has polarization coinciding with the first EP (curves 1,3,5) and with the second EP (curves 2,4,6). $d=35p$. The other parameters are the same as in Fig.2.

Fig. 9 shows the spectra of reflection $R$ (curves 1,2), transmission $T$ (curves 3,4) and absorption A (curves 5,6) in the following two cases: (a) $\text{Im}\varepsilon_1 = 0.05$, $\text{Im}\varepsilon_2 = 0.05$, and (b) $\text{Im}\varepsilon_1 = 0.$, $\text{Im}\varepsilon_2 = 0.1$. The incident light has polarization coinciding with the first EP (curves 1,3,5) and with the second EP (curves 2,4,6). As can be seen from Fig. 9a, in the case of isotropic absorption at exceptional points on the wavelengths $\lambda_5 = \frac{2\pi c}{\omega_5}$ and $\lambda_6 = \frac{2\pi c}{\omega_6}$, the same reflection, transmission, and absorption for incident light with the first and second EPs take place, i.e., reflection transmission and absorption independent of the polarization of the incident light take place at these two wavelengths. As can be seen from Fig. 9b the same take place in the case of anisotropic absorption. However, in this case, narrow transparency bands appear near the wavelengths $\lambda_5$ and $\lambda_6$, i.e., two MIT lines appear here.

And finally, let as consider the peculiarities of reflection, transmission, and absorption near the wavelength $\lambda_2$ (on the wavelength $\lambda_2$ the function g($\lambda$) passes through a minimum with the value g($\lambda$) = 0 at this wavelength). As can be seen from Fig. 1 and Fig. 9 and as our calculations show near this wavelength $\lambda_2$ the reflection, transmittance, and absorption are independent of the polarization of the incident light, moreover both at g>0 (that is when the directions of the incident light and the external magnetic field coincide) and at g<0 (that is when the directions of the incident light and the external magnetic field are oppositely directed), although it is not Dirac point nor exceptional point. And this takes place at any value of the CLC helix pitch, except for those cases when the helix pitch is approximately in the range of (3000÷3500) nm. When the pitch of the helix changes in this interval, then that exceptional lines appear.

### 3. Conclusions

In conclusion, we investigated the properties of CLCs being in external static magnetic field directed along the helix axis. We have shown that in the case of the wavelength dependence of magneto-optic activity parameter, new features appear in the optics of CLCs. Firstly, we investigated the behavior of the function g($\lambda$), and showed that on the local maximum of g($\lambda$) on the wavelength $\lambda_3$ there exist a Dirac point, where new type of PBG is formed. The novelty is that, if the basic PBG (which exist also in the absence of external magnetic field and at its presence undergoes blueshift) is determined by the structure of the CLC, i.e., the chirality sign of the polarization of the incident diffracting light is determined only by the chirality sign of the CLC helix, then for the new one it is determined by the external magnetic field direction (i.e., on whether the directions of the external magnetic field and the incident light are parallel, or they are antiparallel). We showed that at $|g| \geq \varepsilon$ a new (the third) PBG appears in the region $\lambda \leq \lambda_4 = \frac{\pi\sqrt{2\varepsilon_m(1+\sqrt{2})}}{VB_{ext}}$.

Next, we showed that at some values of helix pitch Dirac lines and exceptional lines appears. At some values of the helix pitch in a wide spectral region (about 2000 nm) in the presence of isotropic absorption the coincidence of the real and imaginary parts of two wave vectors occurs. I.e. here we have an exceptional line and here there is a broadband spectral region where there is a reflection, transmission and absorption independent of the polarization of the incident light, moreover, a wide spectral region, where a total absorption takes place independent of the polarization of the incident light. At some other values of the helix pitch in a wide spectral region (about 2000 nm) in the presence of anisotropic absorption the coincidence of the real parts of two wave vectors occurs. Moreover, for one wave vector the imaginary part is equal to zero, while for second one it significantly differs from zero. Hence, here we have the Dirac line. Here wideband CPT takes place for incident light with the first EP and wideband CPA takes place for incident light with the second EP. Moreover, here this system can work as wideband unidirectional transmitter or wideband ideal optical diode.

At large values of the helix pitch, the curves $\text{Re}k_{1z}(\omega)$ and $\text{Re}k_{2z}(\omega)$ intersect at wavelength $\lambda_5 = \frac{2\pi p\sqrt{\varepsilon_m(4\pi^2+V^2B_{ext}^2 p^2)}}{\sqrt{16\pi^4+V^4B_{ext}^4 p^4}}$ and the same curves also touch at wavelength $\lambda_6 = \frac{2\pi p\sqrt{\varepsilon_m}}{\sqrt{4\pi^2+V^2B_{ext}^2 p^2}}$. The curves $\text{Im}k_{1z}(\omega)$ and $\text{Im}k_{2z}(\omega)$ touch at wavelengths $\lambda_5$ and $\lambda_6$ both at isotropic and at anisotropic absorption. Therefore, these points at wavelengths $\lambda_5$ and $\lambda_6$ are exceptional points both at isotropic and at anisotropic absorption. And the reflection transmission and absorption independent of the polarization of the incident light take place at these two wavelengths. However, in the anisotropic absorption case, narrow transparency bands appear near the wavelengths $\lambda_5$ and $\lambda_6$, i.e., two MIT lines appear here.

And finally, at the wavelength $\lambda_2$ (where the function g($\lambda$) passes through a minimum with the value g($\lambda$) = 0) the reflection, transmittance, and absorption are independent of the polarization of the incident light, moreover both at g>0 and at g<0, although it is not Dirac point nor exceptional point.

Let us note, that although our calculations were made at $B_{ext} = 40$ T, however, the observed effects are also observed at much lower values of $B_{ext}$. At smaller values of $B_{ext}$, the frequency range of observation of these effects shift to long wavelengths. The external magnetic field can distort the CLC structure, but since we are considering the case of CLC without local magnetic anisotropy and the external magnetic field directed along helix axis, then according to Mayer's theory [59] we can neglect the helix pitch change in the lowest order approximation. Moreover, the external magnetic field can also directly affect the local components of the dielectric tensor, but since it is quadratic with respect to the external field effect, we can initially neglect it, or consider that the given values for the components of the dielectric permittivity tensor are exactly the values they acquire in an external magnetic field. This does not affect our conclusions, since we do not consider the effects of changing the external magnetic field.

Note also, that the importance of the problem under consideration also lies in the fact that the results are obtained for a relatively simple and long-known structure for which the exact solution of Maxwell's equation is known, and one can obtain simple analytical expressions for the frequencies of Dirac points and exceptional points and establish the causes of the considered effects.

Finally, let us note, that exceptional points and lines have unique properties and are of many applications, such as single-mode lasers [60, 61], unidirectional light propagation [62], coherent absorption [63], enhanced optical sensing [64, 65], laser-absorbers [66–70], topological light steering [71] and others.

**Acknowledgment**

This work was partially supported by the Foundation for the Advancement of Theoretical Physics and Mathematics "BASIS" (Grant No. 21-1-1-6-1).

**Conflict of Interest**

The authors declare no conflict of interest.

**Keywords**

magnetoactive cholesteric liquid crystals, Dirac points and lines, exceptional points and lines, magnetically induced transparency, magnetically induced absorption, coherent total absorption, unidirectional transmission

References

1. F. D. M. Haldane and S. Raghu, Phys. Rev. Lett. **2008**, 100, 013904.
2. Z.Wang, Y. Chong, J. D. Joannopoulos, and M. Soljacic, Nature **2009**, 461, 772.
3. P. A.M. Dirac, Proceedings of the Royal Society A, 117 (1928) 610.
4. A. H. Castro Neto, F. Guinea, N. M. R. Peres, K. S. Novoselov, and A. K. Geim, Rev. Mod. Phys. **2009**, 81, 109.
5. K. S. Novoselov, A. K. Geim, S.V.Morozov et al., Nature, **2005**, 438, 197.
6. M. C. Rechtsman, J. M. Zeuner, A. Tünnermann, S. Nolte, M. Segev, and A. Szameit, Nat. Photon. **2013**, 7, 153.
7. A. B. Khanikaev, S. Hossein Mousavi, W.-K. Tse, M. Kargarian, A. H. MacDonald, and G. Shvets, Nat. Mater. **2013**, 12, 233.

8. M. Hafezi, S. Mittal, J. Fan, A. Migdall, and J. M. Taylor, Nat. Photon. **2013**, 7, 1001.
9. G. Harari, M. A. Bandres, Y. Lumer, M. C. Rechtsman, Y. Chong, M. Khajavikhan, D. N. Christodoulides, and M. Segev, Science **2018**, 359, 4003.
10. B. Zhen, C. W. Hsu, L. Lu, A. D. Stone, M. Soljacic, Phys. Rev. Lett. **2014**, 113, 257401.
11. H. M. Doeleman, F. Monticone, W. d. Hollander, A. Alù, A. F. Koenderink, Nat. Photon. **2018**, 12, 397.
12. J. Gomis-Bresco, D. Artigas, and L. Torner, Nat. Photon. **2017**, 11, 232.
13. F. Bloch, Z. Physik **1929**, 52, 555.
14. C. Zener, Proc. R. Soc. London A **1934**, 145, 523.
15. M. Fleischhauer, A. Imamoglu, J.P. Marangos, Rev. Mod. Phys. **2005**, 77, 633.
16. M. S. Hur, J. S. Wurtele, G. Shvets. Phys. Plasmas. **2003**, 10, 3004.
17. B. Zhen, C.W. Hsu, Y. Igarashi, Nature **2015**, 525, 354.
18. K.-H. Kim, M.-S. Hwang, H.-R. Kim, et al. Nat. Commun. **2016**, 7, 13893.
19. Ş.K. Özdemir, S. Rotter, F. Nori, L. Yang. Nat. Mat. **2019**, 18, 783.
20. M.-.A. Miri, A. Alù. Science, **2019**, 363, 42.
21. J. Gomis-Bresco, D. Artigas, L. Torner. Phys. Rev. Res. **2019**, 1, 033010.
22. M. Parto, Y.G. N. Liu, B. Bahari, et al. Nanophotonics **2021**, 10, 403.
23. A. Li, H. Wei, M. Cotrufo, et al. Nat. Nanotech. **2023**, 18, 706.
24. M. Król, I. Septembre, P. Oliwa, et al. Nat. Commun. **2022**, 13, 5340.
25. T. Wang. New J. Phys., **2022**, 24, 113016.
26. J.H.D. Rivero, L. Feng, L. Ge. Phys. Rev. B **2023**, 107, 104106.
27. D. Anderson, M. Shah, L. Fan. Phys. Rev. Appl. **2023**, 19, 034059.
28. A. Lezama, S. Barreiro, A.M. Akulshin. Phys. Rev. A **1999**, 59, 4732.
29. L. Feng, R. El-Ganainy, and L. Ge, Nat. Photon. **2017**, 11, 752.
30. R. Gad, J.G. Leopold, A. Fisher, F.R. Fredkin, A. Ron. Phys. Rev. Lett. **2012**, 108, 155003.
31. J.H. Yan, P. Liu, et al., Nat. Commun. **2015**, 6, 7042.
32. D. Floess, M. Hentschel, et al., Phys. Rev. X **2017**, 7, 021048.
33. A. Christofi, Y. Kawaguchi, A. Alù, A.B. Khanikaev. Opt. Lett. **2018**, 43, 1838.
34. Z. Wu, J. Li, Y. Wu. Phys. Rev. A **2022**, 106, 053525.
35. M. N. Winchester, M. A. Norcia, et al., Phys. Rev. Lett. **2017**, 118, 263601.
36. A. Sommer, Physics **2017,** 10, 70.
37. X. Zhang, C.L. Zou, L. Jiang, H.X. Tang, Phys. Rev. Lett. **2014**, 113, 156401.
38. G.H. Dong, D.Z. Xu, P. Zhang, Phys. Rev. A **2020**, 102, 033717.
39. K. Ullah, M. T. Naseem, Ö. E. Müstecaplıoğlu. Phys. Rev. A **2020**, 102, 033721.
40. M. Mitov, Adv. Mater. **2012**, 24, 6260.
41. A.H. Gevorgyan, Mol. Cryst. Liq. Cryst. **2002**, 382, 1.
42. I. Bita, E.L. Thomas, J. Opt. Soc. Amer. A., **2005**, 22, 1199.
43. I.-C. Khoo, Liquid Crystals. 2nd edn. J. Wiley & Sons, Hoboken, New Jersey (2007).
44. A.H. Gevorgyan. Opt. Mat., **2021**, 113, 110807.
45. A.H. Gevorgyan, S.S. Golik, Y.N. Kulchin, N.A. Vanyushkin. Optik **2021**, 247, 167927.
46. A.H. Gevorgyan, S.S. Golik. Opt. Las. Techn. **2022**, 149, 107810.
47. A.H. Gevorgyan. J. Appl. Phys. **2022**, 132, 133101.
48. P. K. Challa, V. Borshch, et al., Phys. Rev. E **2014**, 89, 060501(R).
49. S. M. Salili, J. Xiang, et al., Phys. Rev. E **2016**, 94, 042705.
50. J.M. Caridad, C. Tserkezis, et al., Phys. Rev. Lett. **2021**, 126, 177401.
51. A.H. Gevorgyan, S.S. Golik, et al., Materials **2021**, 14, 2172.
52. A.H.Gevorgyan. Opt. Lett. **2021**, 46, 3616.
53. A.H. Gevorgyan. J. Mol. Liq. **2021**, 339, 117276.
54. A.H. Gevorgyan. Phys. Rev. E **2022**, 105, 014701.
55. A.H.Gevorgyan. Phys. Lett. A **2022**, 443, 128222.
56. A.H. Gevorgyan, N.A. Vanyushkin, et al. J. Opt. Soc. Amer. B **2023**, 40, 1986.
57. H.L. De Vries. Acta Crystallographica **1951**, 4, 219.

58. E.I. Kats. Sov. Phys. JETP, **1971**, 32, 1004.
59. R.B. Meyer, Appl. Phys. Lett., **1968,** 12, 281.
60. H. Hodaei, M. A. Miri, M. Heinrich, D. N. Christodoulides and M. Khajavikhan, Science **2014**, 346, 975.
61. F. Liang, J. W. Zi, R.M. Ma, Y. Wang and X. Zhang, Science **2014,** 346, 972.
62. L. Feng, Y. L. Xu, W.S. Fegadolli, M.H. Lu, J. Oliveira, V. R. Almeida, Y. F. Chen and A. Scherer, Nat. Mater. **2013**, 12, 108.
63. Y. Sun, W. Tan, H. Q. Li, J. Li and H. Chen, Phys. Rev. Lett. **2014**, 112, 143903.
64. J. Wiersig, Phys. Rev. Lett. **2014**, 112, 203901.
65. Z.-P. Liu et al, Phys. Rev. Lett. **2016**, 117, 110802.
66. S. Longhi, Phys. Rev. A **2010**, 82, 031801(R).
67. Y. D. Chong, G. Li and A. D. Stone, Phys. Rev.Lett. **2011**, 106, 093902.
68. S. Longhi and L. Feng, Opt. Lett. **2014**, 39, 5026.
69. P. A. Kalozoumis, C. V. Morfonios, G. Kodaxis, F. K. Diakonos and P. Schmelcher, Appl. Phys. Lett. **2017**, 110, 121106.
70. V. Achilleos, V. Aur´egan and V. Pagneux, Phys. Rev. Lett. **2017**, 119, 243904.
71. H. Zhao, X. Qiao, T. Wu, B. Midya, S. Longhi and L. Feng, Science **2019**, 365, 1163.